\pdfoutput=1

\documentclass[review]{elsarticle}

\usepackage{graphicx}
\usepackage{amssymb}
\usepackage{tabularx}
\usepackage{siunitx}
\usepackage{placeins}
\usepackage{color}
\usepackage{soul}

\biboptions{numbers}
\biboptions{sort&compress}

\journal{Applied Surface Science}

\begin{document}

\begin{frontmatter}

\title{Protein arrangement on modified diamond-like carbon surfaces - an ARXPS study}

\author[label1]{Reece N. Oosterbeek \corref{cor1} }
\cortext[cor1]{Corresponding author}
\ead{reece.oosterbeek@auckland.ac.nz}
\author[label2]{Christopher K. Seal}
\author[label1]{Margaret M. Hyland}

\address[label1]{Department of Chemical and Materials Engineering, The University of Auckland, Private Bag 92019, New Zealand}
\address[label2]{Light Metals Research Centre, The University of Auckland, Private Bag 92019, New Zealand}

\begin{abstract}
Understanding the nature of the interface between a biomaterial implant and the biological fluid is an essential step towards creating improved implant materials. This study examined a Diamond-like Carbon coating biomaterial, the surface energy of which was modified by Ar\textsuperscript{+} ion sputtering and laser graphitisation. The arrangement of proteins was analysed by angle resolved X-Ray photoelectron spectroscopy, and the effects of the polar component of surface energy on this arrangement were observed. It was seen that polar groups (such as CN,CO) are more attracted to the coating surface due to the stronger polar interactions. This results in a segregation of these groups to the DLC-protein interface; at increasing takeoff angle (further from to DLC-protein interface) fewer of these polar groups are seen. Correspondingly, groups that interact mainly by dispersive forces (CC, CH) were found to increase in intensity as takeoff angle increased, indicating they are segregated away from the DLC-protein interface. The magnitude of the segregation was seen to increase with increasing polar surface energy, this was attributed to an increased net attraction between the solid surface and polar groups at higher polar surface energy ($ \gamma^{p}_{S} $).
\end{abstract}


\end{frontmatter}

\section{Introduction}
The use of medical implants drives the need understand the interactions that are taking place at the  interface between biological fluid and implant materials. Diamond-like Carbon (DLC) coatings are commonly applied as coatings on arterial stents \cite{Robertson02}, and in this application in particular, the interactions at the interface are of key importance. This is due to adsorption of proteins - a commonplace phenomenon that occurs when an implant material is introduced, and in some cases this can lead to thrombus formation and embolism (formation of blood clots that can block blood vessels and cause tissue death) \cite{Lousinian07}. Understanding the nature of the interface and the initial protein layer is therefore an important step towards developing superior and safer implant materials.\\

The adsorption of BSA  (Bovine Serum Albumen) onto stainless steel surfaces has been widely studied \cite{Zanna06, Gruian12, Hedberg12}, with results showing significant interactions between the protein and the surface. These interactions result in conformational and structural changes of the protein, and even migration of surface species. There have been few XPS  (X-Ray Photoelectron Spectroscopy) studies on protein adsorption onto DLC surfaces, however Ahmed et al. \cite{Ahmed13} found that increased Si doping increases sp\textsuperscript{3} content, as well as decreasing adsorption. These results suggest that surface chemistry has a strong effect on both the overall adsorption, as well as the structure of the adsorbed layer. In addition, it is known that upon adsorption, protein molecules are likely to undergo conformational changes and denaturation \cite{SilvaBermudez13}, however what these changes are has not yet been identified for DLC coatings.\\

The polar component of surface energy appears to be the major contributor to the amount of protein adsorption \cite{Feng03, Yang09, SilvaBermudez13}. However, the arrangement of this initial protein layer, and the effects of surface energy on this, has not been comprehensively studied. The lack of information in this area appears to be a common trend in the study of protein adsorption to surfaces, with Silva-Bermudez and Rodil identifying that more work need to be done to correlate the properties of the surface to protein adsorption mechanisms \cite{SilvaBermudez13}. Studies by Browne et al. and Baio et al. \cite{Browne04, Baio11} of protein adsorption onto polystyrene and calcium phosphate surfaces shows that adsorption of protein can be significantly affected by substrate surface chemistry, and it is suggested that the changes in adsorption behaviour seen require a change in conformation of the protein molecule. Feng et al. \cite{Feng03} found that protein adsorption depended mainly on the polar component of surface energy and the hydroxyl groups, but there has been no investigation into whether this interaction causes a change in the arrangement of the adsorbed protein. In the present study, the arrangement of different protein functional groups adsorbed onto a DLC surface was investigated, and the effect of the polar surface energy component on this arrangement was analysed.\\

Studies on protein adsorption on DLCs have also shown that the sp\textsuperscript{2}/sp\textsuperscript{3} bonding ratio in the DLC material has a significant effect on the haemocompatibility of the surface, with increasing sp\textsuperscript{3} content increasing adsorption of Human Serum Albumin relative to Fibrinogen \cite{Logothetidis07}. Therefore in this study the sp\textsuperscript{2}/sp\textsuperscript{3} bonding ratio was altered by surface treatments (specifically Ar\textsuperscript{+} ion sputtering and laser graphitisation) in order to observe the effect on protein adsorption.\\

In order to examine the arrangement of proteins on the biomaterial surface, a protein layer was adsorbed onto the surface of the modified and unmodified DLC coatings. BSA was used as it is often used as a model for protein adsorption from blood \cite{Lee88}, as well as being a well-characterised protein that is often used to study protein interactions with solid surfaces \cite{Kwoka05}. This allows the present study to be relevant for applications involving interactions of blood with DLCs such as arterial stents. 

\section{Materials and Methods}

	\subsection{Diamond-like Carbons}
Diamond-like carbons (DLCs) are a form of amorphous carbon, with a mixture of both sp\textsuperscript{2} and sp\textsuperscript{3} bonded carbon. The properties of the DLC coating are heavily dependent on this sp\textsuperscript{2} to sp\textsuperscript{3} bonding ratio \cite{Lifshitz99}. In this work, Ar\textsuperscript{+} ion sputtering and laser graphitisation are used to graphitise the surface of the DLC coatings, that is, to convert some of the sp\textsuperscript{3} bonded carbon to sp\textsuperscript{2}; the form found in graphite.\\

The diamond-like carbon coating used was a DLC coating with metallic interlayers to improve adhesion to the substrate, and was deposited by a combination of PACVD (Plasma Assisted Chemical Vapour Deposition) and PVD (Physical Vapour Deposition) methods.

	\subsection{Ar\textsuperscript{+} Ion Sputtering}
Ar\textsuperscript{+} ion sputtering was carried out to induce surface graphitisation of DLC coatings, using the method developed by Viljoen et. al. \cite{Viljoen96}. The ion beam used was the Kratos Minibeam III (an accessory technique to the Kratos Axis ultraDLD XPS - see below). Samples were prepared in the same way as for XPS. For ion sputtering, the chamber pressure of the XPS was slightly higher than usual (1.5x10\textsuperscript{-7} Torr), due to the presence of the sputtering gas.  The ion beam energy was 5.2 keV  with an emission current of 6mA. The raster size was adjusted to 3x3mm, and the ion beam turned on for 4 minutes. In order to graphitise a larger area, the raster square was moved to several different positions on the sample rather than increasing the area (which would give a lower emission current density, seen by trials to result in non-optimal graphitisation).\\

To ensure that any changes in surface chemistry seen were a result of actual changes in the DLC coating (rather than just removal of adsorbed species), the sputtered DLC was exposed to atmosphere for 10 minutes, before being re-analysed. If the changes seen are simply due to removal of adsorbed species, we would expect to see the C1s peak shift back to the higher binding energy associated with sp\textsuperscript{3} carbon after exposure to air, as adsorbed species (adventitious carbon) are usually of the sp\textsuperscript{3} configuration \cite{Moulder92}. However if an actual change in surface chemistry is achieved there should be little change in the C1s peak after exposure to air - it should remain at the lower binding energy associated with sp\textsuperscript{2} (graphitised) carbon.

	\subsection{Laser Graphitisation}
Laser graphitisation experiments were carried out using a femtosecond laser micromachining system based on a Ti:Sapphire amplified femtosecond laser (Mantis (oscillator) and Legend Elite (amplifier), Coherent Inc., USA); similar to the method used by Dumitru et. al. \cite{Dumitru03}. This laser supplies 800 nm wavelength, ultrashort (100 fs) pulses at a repetition rate of 500Hz to a JPSA IX-100 micromachining platform under atmospheric conditions. The laser was set up to produce a square spot (50x\SI{50}{\micro\metre}) with an energy density of 1.28 J/cm\textsuperscript{2}. Beam attenuation was set to 69\% (using a waveplate and polariser variable attenuator) to give the desired energy density of 0.4 J/cm\textsuperscript{2} to compare with the work of Dumitru et. al. The micromachining platform was programmed to scan and fire across the whole sample surface, resulting in a uniformly laser treated surface.

	\subsection{Protein Adsorption}
A 1mg/mL BSA solution was prepared by dissolving BSA powder into distilled water. The solution was be prepared immediately before the adsorption test, and then analysed without delay, to prevent degradation of the protein. The sample was immersed in 20 mL of BSA solution in a plastic vial, unagitated at room temperature, for 60 minutes. After immersion the sample was rinsed with distilled water to remove any solution that may have dried on the surface (not adsorbed).

	\subsection{Analysis Methods}
XPS was carried out on a  Kratos Axis ultraDLD, using a monochromated Al K$\alpha$ X-Ray source with a power of 150 W. The chamber pressure was approximately 1.5x10\textsuperscript{-7} Torr. Survey scans were performed at pass energy of 160 eV from 0 to 1350 eV, while core level scans were carried out at 20 eV to improve resolution. Angle Resolved XPS was used to alter the analysis depth for analysis of protein arrangment - a higher take-off angle results in a smaller analysis depth. Take-off angles of 0, 20, 40, 60, and 80$^{\circ}$ were used to produce a depth profile of the surface (take-off angle is defined here as the angle between the collection angle and surface normal).\\

Peak fitting was carried out using CasaXPS software, fitting with a Shirley-type background and Gauss-Lorentz component peak shapes. Graphitic (sp\textsuperscript{2}) carbon was fitted with the Doniach-Sunjic asymmetry known to be characteristic of graphite \cite{Speranza06}. Components were fitted so that peak widths remained similar for all components. The sp\textsuperscript{3} and sp\textsuperscript{2} components of the DLC were fitted using two peaks at 285.4 eV and 284.8 eV respectively \cite{Robertson02, Moulder92}.\\

The components of the adsorbed protein were determined from analysis of a BSA standard sample. Samples were cleaned with isopropanol and air dried before analysis to remove surface contaminants (for samples with adsorbed proteins distilled water was used instead), and six repeats were carried out for all samples to ensure reliability and repeatability of data. Charge neutralisation was used for the BSA protein standard, which was prepared by evaporating BSA solution (1mg/mL) in a gold-coated sample stub. The sp\textsuperscript{2} carbon component at 284.8 eV was used as a binding energy reference.\\

Scanning Electron Microscopy was carried out using a Phillips FEI XL30S, with a Field Emission Gun (FEG) electron source. The imaging parameters such as accelerating voltage are seen in the information bar on the image.\\

The surface energy of treated and untreated DLCs were calculated from measurements of contact angles of di - iodomethane and water, a polar and non-polar fluid.  Prior to measurement, samples were cleaned with ethanol and distilled water. The instrument used was a KSV CAM 101 goniometer, using the sessile drop technique, where a drop of the probe liquid is dropped onto the surface and the contact angle of this drop with the surface is measured. This measurement was done by the CAM 2008 software. Between measurements, samples were rinsed with distilled water to remove any residual fluid.\\

The surface energy was calculated by the Owens and Wendt equation \cite{Owens69}, where the contact angle with two different fluids is required. The Owens and Wendt equation is shown below, where $\gamma$ is the surface energy, $\theta$ is the contact angle, the subscripts \textit{S} and \textit{L} denote the solid and liquid parts of the system respectively, and the superscripts \textit{p} and \textit{d} denote the polar and dispersive components of surface energy respectively.
\begin{equation}
\gamma_{L}(1+\cos\theta) = 2\sqrt{\gamma_{S}^{d}\gamma_{L}^{d}} + 2\sqrt{\gamma_{S}^{p}\gamma_{L}^{p}}
\end{equation}

\section{Results and Discussion}

	\subsection{Surface Chemistry}
\begin{figure}
\centering
\includegraphics[width=0.5\columnwidth]{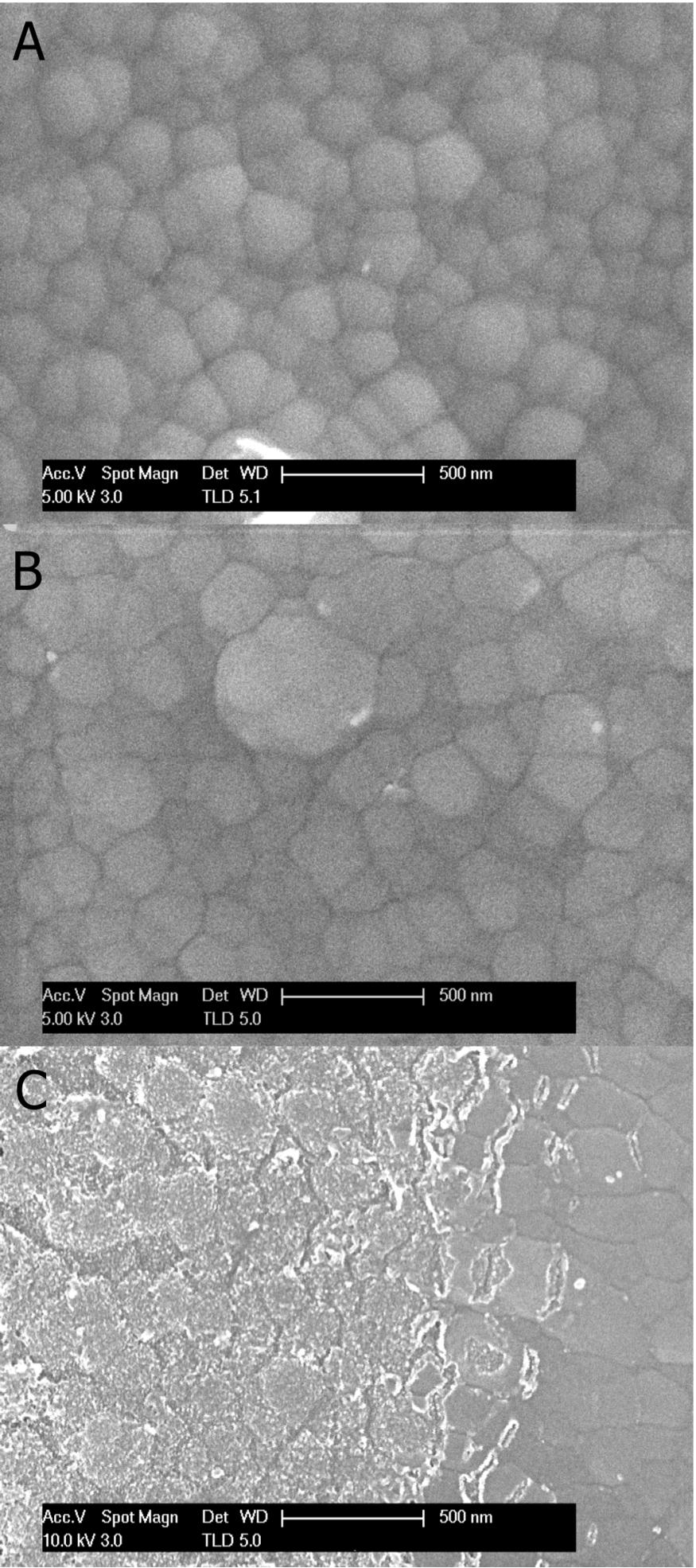}
\textbf{\caption{\label{fig:Fig_1}SEM images of (a) the standard DLC, (b) Ar\textsuperscript{+} sputtered DLC and (c) laser graphitised DLC.}}
\end{figure}

A summary of surface chemistry components measured by XPS is shown in Table \ref{tab:surfacechem}. The unmodified DLC is primarily composed of C-C bonds (sp\textsuperscript{3} and sp\textsuperscript{2}), with some CO and C=O bonds also found. After Ar\textsuperscript{+} ion sputtering, no change was seen in the SEM image from the unmodified DLC (Figure \ref{fig:Fig_1}), indicating that this graphitisation had not affected the surface morphology, but only the surface chemistry. The nodular structure seen in these images is the result of the PACVD and PVD processes used to produce the DLC coating.\\

\begin{figure}
\centering
\includegraphics{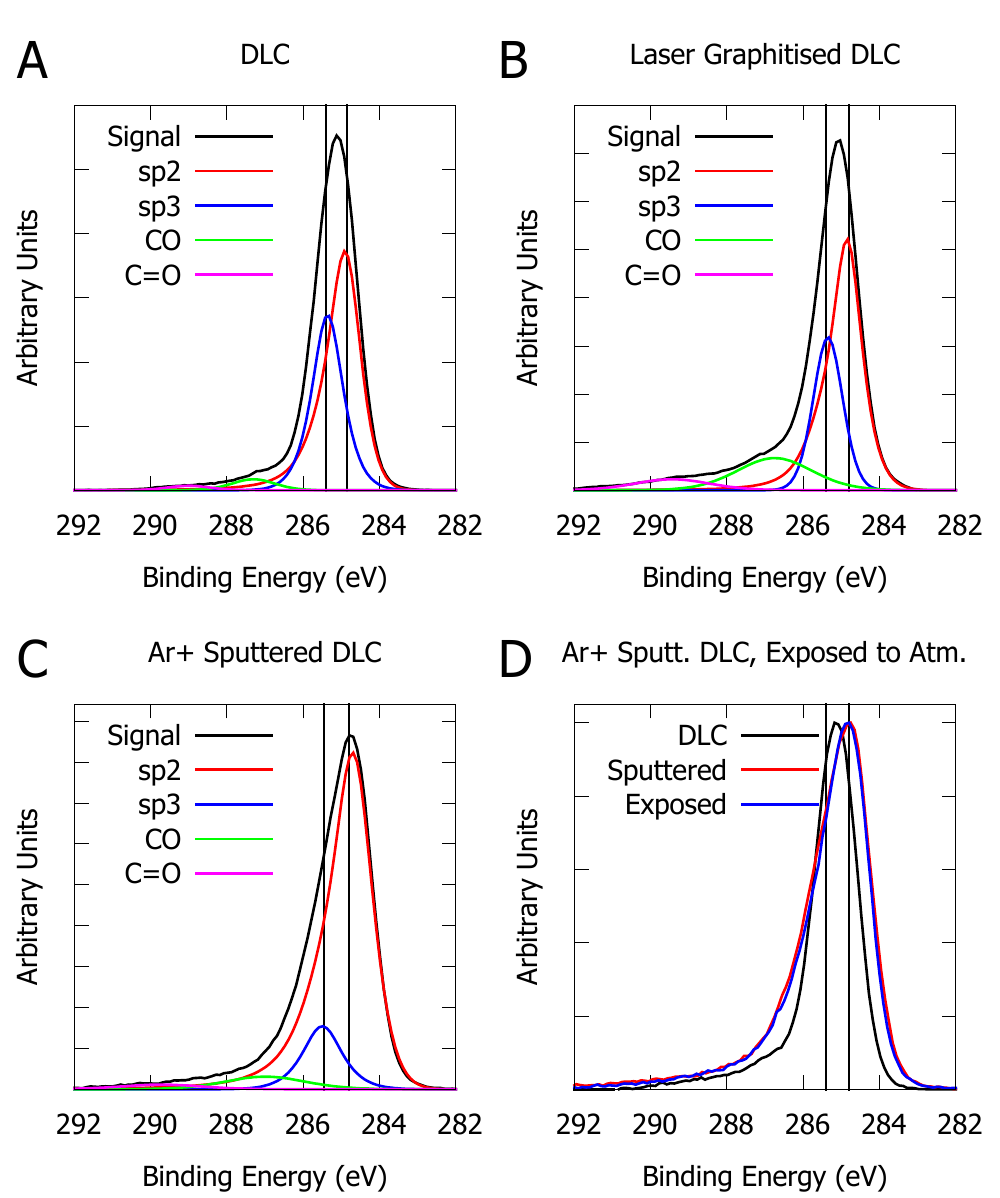}
\textbf{\caption{\label{fig:Fig_2}XPS C1s spectra for (a) the standard DLC, (b) laser graphitised DLC, (c) Ar\textsuperscript{+} sputtered DLC, and (d) Ar\textsuperscript{+} sputtered DLC before and after exposure to atmosphere. Vertical lines show the positions of the sp\textsuperscript{3} and sp\textsuperscript{2} components at 285.4 and 284.8eV respectively.}}
\end{figure}

A significant change is seen in the XPS C1s spectrum after Ar\textsuperscript{+} ion sputtering, as seen in Figure \ref{fig:Fig_2}A,C,D. The main peak shifts to the right (lower binding energy), indicating a decrease in the sp\textsuperscript{3}/sp\textsuperscript{2} fraction from 0.77 to 0.19 (as the sp\textsuperscript{2} component has lower binding energy than sp\textsuperscript{3}).  This can be explained by the mechanism proposed by Briand et. al. \cite{Briand09}. The positively charged Ar\textsuperscript{+} ion penetrates into the surface and extracts electrons from the material, creating positive holes. On dielectric materials like DLCs, these positive holes cannot be neutralised immediately, creating high local stresses, in turn causing surface modification (graphitisation). The energy (in the form of moving ions) input into the material allows C atoms to transform to the stable graphite (sp\textsuperscript{2}) structure from meta-stable (at ambient conditions) diamond sp\textsuperscript{3}. \\

The SEM image of the laser graphitised DLC (Figure \ref{fig:Fig_1}) shows how the laser square has been rastered across the surface, causing a surface modification. The modification is mainly graphitisation of the DLC, seen by the decrease in sp\textsuperscript{3}/sp\textsuperscript{2} ratio, but some oxidation of the surface has occured, as shown by Figure \ref{fig:Fig_2}B - increased CO and C=O components are found, as well as the sp\textsuperscript{2} and sp\textsuperscript{3} components of the DLC. This is the result of the high temperatures generated at the surface by the laser.\\

In order to ensure that any changes in surface chemistry of the Ar\textsuperscript{+} sputtered DLC were a result of changes in the DLC coating rather just than removal of adsorbed species, samples were exposed to atmosphere for 10 minutes, before being re-analysed by XPS. From the XPS spectra in Figure \ref{fig:Fig_2}D, it is easy to see the shift in binding energy associated with graphitisation. Comparing the two spectra before and after exposure to air, there is essentially no difference, indicating that the binding energy shift from 285.4 eV to 284.8 eV (i.e. graphitisation) seen is a result of a chemical change, rather than simply removal of adsorbed species (as adsorbed species would be manifested as hydrocarbon species at 285.4 eV\cite{Moulder92}). This is verified by quantifications of the various components of these peaks, giving an sp\textsuperscript{3}/sp\textsuperscript{2} ratio of 0.77 $(\pm0.01)$ before sputtering, 0.19 $(\pm0.02)$ after sputtering, and 0.17 $(\pm0.01)$ after exposure to air. From survey scans we are able to quantify the oxygen composition: 6.9$\% (\pm0.4)$ for the DLC, 0.6$\% (\pm0.1)$ after sputtering, and 2.5$\% (\pm0.2)$ after exposure to air. This does suggest that some adsorption has taken place, however because our analysis focuses on the C1s peak, we are most concerned with the effects on this. Figure \ref{fig:Fig_2}D shows significant change after sputtering, and no change after exposure of the sputtered sample to air, so we can reasonably conclude that the changes seen are not a result of removal of adsorbed species.\\

Both the unmodified DLC and the Ar\textsuperscript{+} ion sputtered DLC are seen to have a dispersive component of surface energy that is significantly greater than the polar component (see Table \ref{tab:surface-energy}),indicating that the surface is composed primarily of non-polar bonds (such as C-C bonds). This agrees with the composition already measured by XPS, where only a small polar component (5.4 - 5.5$\%$ CO and C=O bonding for both) was found. Ar\textsuperscript{+} ion sputtering did not significantly alter the dispersive surface energy component, however the polar surface energy decreases from 5.7 to 3.5 mN/m, which can only have been caused by the change in sp\textsuperscript{3}/sp\textsuperscript{2} ratio.\\

Laser graphitisation appreciably changes the surface energy of the DLC. The most notable change is the increase in the polar surface energy component from 5.7 to 13.6 mN/m, due to the oxidation (which contains polar CO and C=O bonds) caused by laser graphitisation.\\

\begin{figure}
\centering
\includegraphics[width=0.5\columnwidth]{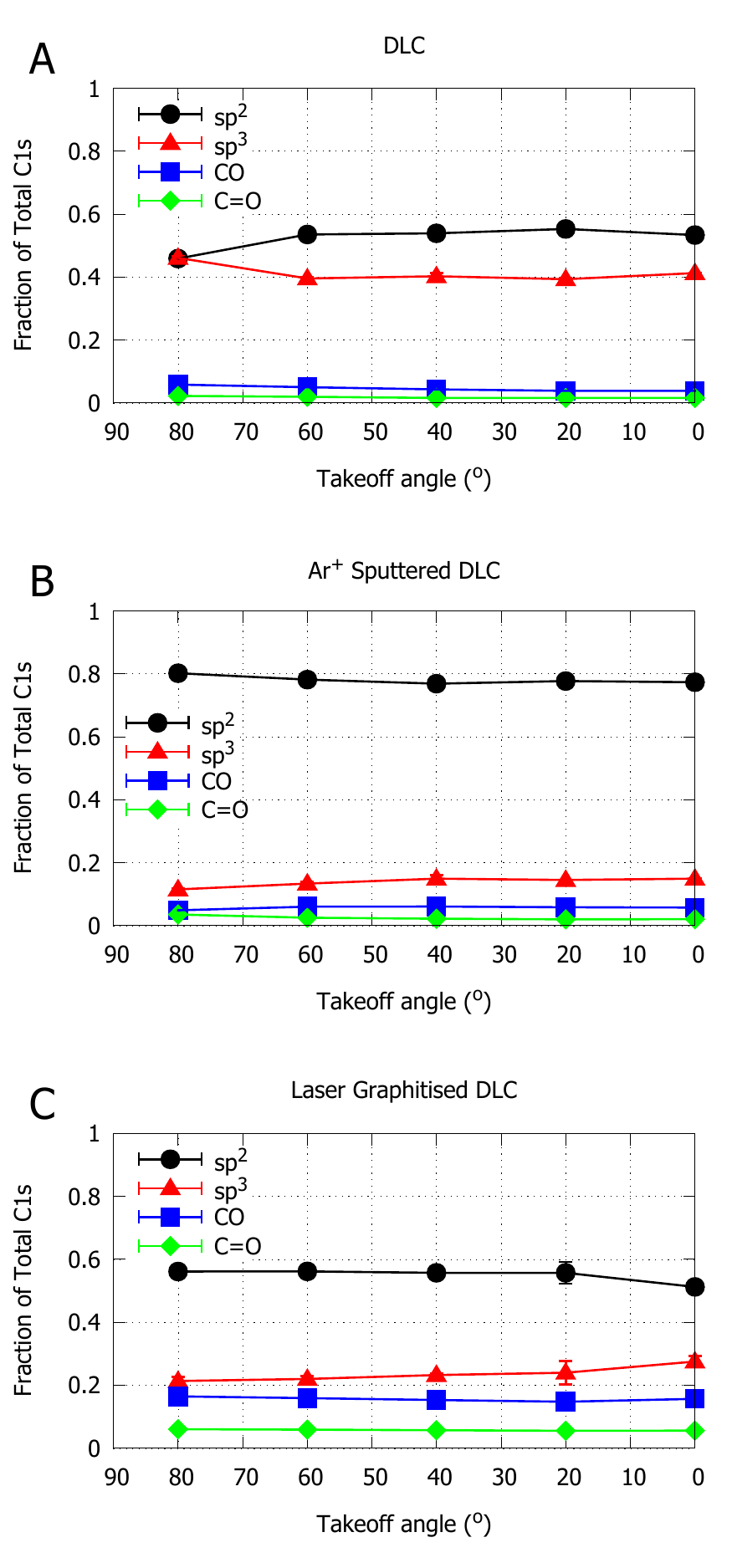}
\textbf{\caption{\label{fig:Fig_3}Graphs showing the variation in C1s groups (sp\textsuperscript{2}, sp\textsuperscript{3}, CO and C=O) as a function of takeoff angle for (a) unmodified DLC, (b) Ar\textsuperscript{+} sputtered DLC, and (c) laser graphitised DLC. Higher takeoff angle indicates greater surface sensitivity.}}
\end{figure}

Angle Resolved XPS (ARXPS) was used to characterise the DLC surfaces as a function of depth, before BSA adsorption. Figure \ref{fig:Fig_3} shows the arrangement of different groups in the DLC coating as a function of takeoff angle. The unmodified DLC shows negligible change in the CO and C=O groups with depth, however at higher takeoff angles (i.e. more surface sensitive) the sp\textsuperscript{3} content decreases slightly. After Ar\textsuperscript{+} ion sputtering sp\textsuperscript{3} content follows the same trend as for the unmodified DLC, albiet at a lower sp\textsuperscript{3} content to begin with. The laser graphitised DLC shows an increasing trend for sp\textsuperscript{3}, CO and C=O groups with increasing takeoff angle. This trend is uncertain however, as SPM (Scanning Probe Microscopy) showed that laser graphitisation resulted in an increased surface roughness, with a variation in the z-direction of $>$200nm, hence it cannot be verified that this is an actual trend rather than a surface roughness effect. Because of the uncertainty introduced by this surface roughness, we have removed the laser graphitised DLC from our analysis from this point forward.

\subsection{Protein Overlayer}
\begin{figure}
\centering
\includegraphics{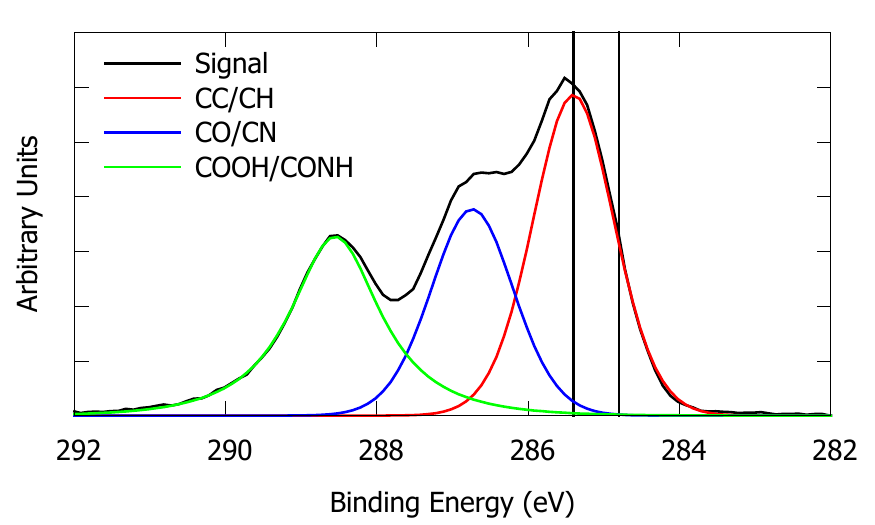}
\textbf{\caption{\label{fig:Fig_4}XPS spectrum, showing the three components of the BSA C1s peak. Vertical lines show the positions of the sp\textsuperscript{3} and sp\textsuperscript{2} components at 285.4 and 284.8eV respectively.}}
\end{figure}

To establish the functional groups present and their relative proportions, a standard sample with BSA adsorbed onto a gold-coated sample stub was analysed. The C1s spectrum was fitted with three peaks corresponding to common protein functional groups: CC/CH at 285.4eV (equivalent to sp\textsuperscript{3} bonded carbon), CO/CN at 286.7eV, and COOH/CONH at 288.6eV. These are consistent with literature \cite{Moulder92, Zanna06} and are shown in Table \ref{tab:BSA-standard} and Figure \ref{fig:Fig_4}.\\

From the XPS spectra of BSA adsorbed onto DLC surfaces, four different components could be deconvoluted from within the C1s envelope: 
\begin{itemize}
\item sp\textsuperscript{2} at 284.8 eV, from the DLC only
\item sp\textsuperscript{3}/CC/CH at 285.4 eV, from the DLC and CC/CH bonds in the protein
\item CO/CN at 286.7 eV, from CO bonds in the DLC and CO/CN bonds in the protein
\item C=O/COOH/CONH at 288.6 eV, from C=O bonds in the DLC and COOH/CONH bonds in the protein
\end{itemize}
The relative proportions of the different functional groups are consistent with published data on the relative proportions of these groups in the bulk protein \cite{Peters85, Pradier03}. This indicates that the protein on the gold surface has not segregated and the values shown in Table \ref{tab:BSA-standard} can be taken as indicative of the bulk BSA.\\

In order to analyse the protein arrangement, the XPS signal from the BSA was deconvoluted from the combined signal that contains both the BSA overlayer and the underlying DLC. The depth profiles of the modified and unmodified DLCs (before BSA adsorption) were used to characterise the amounts of different chemical groups (i.e. sp\textsuperscript{3}, CO and C=O) relative to sp\textsuperscript{2}. The sp\textsuperscript{2} carbon was used as a reference as it is the only carbon component that is present in the DLC but not in the protein overlayer. From these depth profiles it was possible to calculate the amount of sp\textsuperscript{3}, CO, and C=O carbon relative to sp\textsuperscript{2} for each DLC as a function of analysis depth (in terms of inelastic mean free path $\lambda_{IMFP}$).\\

The depth of the protein overlayer (assuming formation of a homogeneous overlayer) was then estimated using the Strohmeier equation, where \textit{N} represents the volume density of atoms, \textit{I} represents the intensity (peak area), and the subscripts \textit{s} and \textit{o} indicate the substrate and overlayer respectively:
\begin{equation}
d = \lambda_{o}\cos\Theta\ln\left[ \frac{N_{s}\lambda_{s}I_{o}}{N_{o}\lambda_{o}I_{s}}+1 \right] 
\end{equation}

For each of the DLC surfaces analysed, the protein overlayer thickness (see Table \ref{tab:surfacechem}) was found to be of the order of the inelastic mean free path (48.32\AA{} \cite{Tan06}). These values are similar the 35\AA{} $\pm8$ measured by Frateur et al. \cite{Frateur07} by EQCM. The protein layer adsorbed onto the untreated DLC surface is significantly thicker than that on the Ar\textsuperscript{+} ion sputtered DLC though, which is thought to be a result of the larger polar surface energy component (see Table \ref{tab:surface-energy}) increasing the bonding between substrate and protein, as seen in multiple other studies \cite{Feng03, Yang09, SilvaBermudez13}. The BSA protein itself has approximate dimensions of a 40 x 40 x 140{\AA} spheroid, \cite{Peters85, Peters95, McClellan03} therefore these thickness measurements indicate that adsorption has taken place in the form of a monolayer, oriented side-on (long axis parallel to the surface).\\

\begin{figure}
\centering
\includegraphics{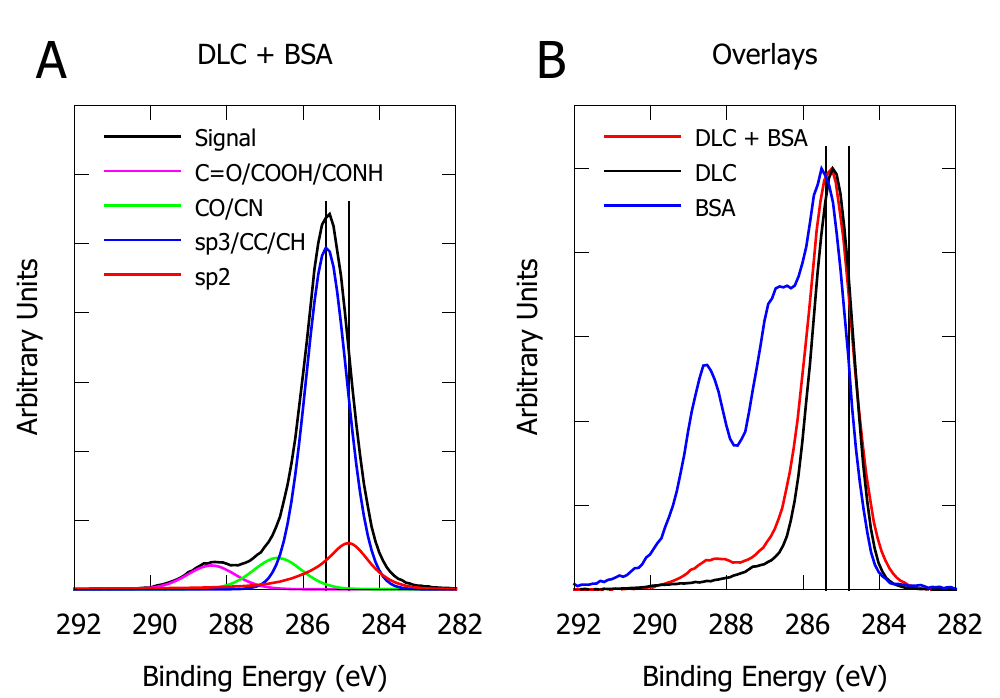}
\textbf{\caption{\label{fig:Fig_5}(a) XPS C1s spectrum of the DLC with BSA overlayer, showing the various DLC and BSA components. (b) Overlays of the C1s spectrum for the DLC, DLC with adsorbed BSA, and BSA only. Vertical lines show the positions of the sp\textsuperscript{3} and sp\textsuperscript{2} components at 285.4 and 284.8eV respectively.}}
\end{figure}

The fact that the layer thickness and inelestic mean free path are of similar magnitude means that the signal from the underlying DLC will originate from the very near surface of the DLC (i.e. close to the DLC-protein interface) and have the corresponding composition as measured by depth profiling of the DLC surface before protein adsorption. This known composition of the DLC (in terms of the amounts of sp\textsuperscript{3}, CO and C=O  carbon relative to sp\textsuperscript{2}) was subtracted from the signal obtained (again normalised to sp\textsuperscript{2}), to determine the signal from the BSA alone. Shown in Figure \ref{fig:Fig_5} is an example of the XPS C1s spectrum of BSA adsorbed onto one of the DLC surfaces. Overlays of the DLC, DLC with adsorbed BSA, and BSA only C1s spectrums are also shown here, to demonstrate the change seen on adsorption of BSA, and how this relates to the composition of the pure BSA.\\

\begin{figure}
\centering
\includegraphics[width=0.5\columnwidth]{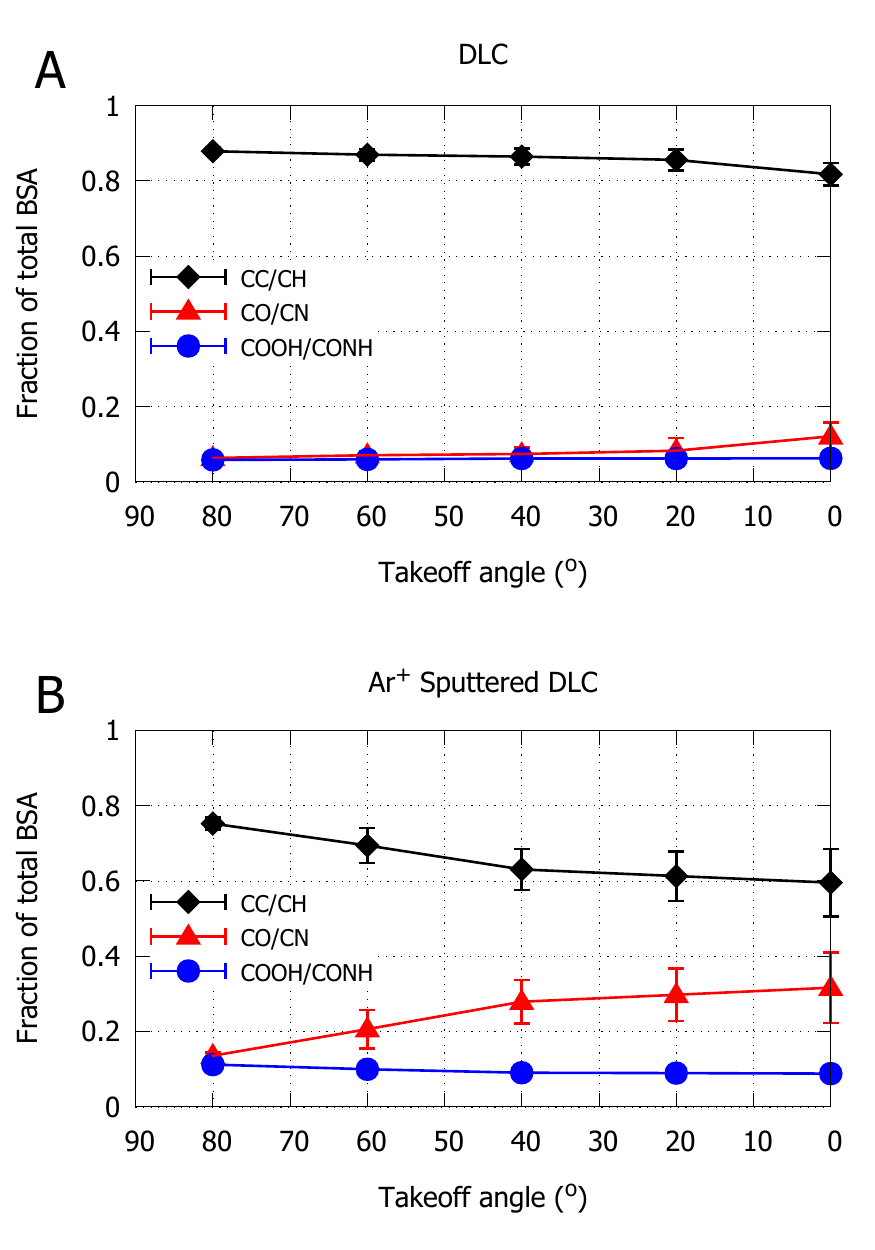}
\textbf{\caption{\label{fig:Fig_6}Protein arrangement graphs, showing the composition of the BSA overlayer as a function of takeoff angle on (a) unmodified DLC and (b) Ar\textsuperscript{+} ion sputtered DLC. Higher takeoff angle indicates greater surface sensitivity.}}
\end{figure}

After subtracting the signal from the underlying DLC, the arrangment of the adsorbed protein overlayer could be analysed - this is shown in the protein arrangement graphs (Figure \ref{fig:Fig_6}) showing the composition of the BSA overlayer as a function  of depth. Both the unmodified DLC and the Ar\textsuperscript{+} ion sputtered DLC display a greater proportion of sp\textsuperscript{3} carbon than the bulk BSA (42\%, see Table \ref{tab:BSA-standard}), which indicates that these CC/CH bonds are facing outwards - away from the substrate-protein interface.\\

For the unmodified DLC, the relative amount of CO/CN carbon increases (with a corresponding decrease in the CC/CH group) as the analysis depth increases, indicating that the CO/CN groups are slightly segregated towards the protein - DLC interface. This is consistent with the trend seen by Premathilaka \cite{Sashini} in ARXPS studies of whey protein adsorption, even though these use a different protein at a higher temperature of 7$^{\circ}$C. This same trend is observed in the Ar\textsuperscript{+} ion sputtered DLC, which shows a much larger change in relative amounts of the CC/CH and CO/CN groups with depth - CC/CH decreases with increasing analysis depth, while the CO/CN group increases. One notable difference between the protein adsorbed onto the modified and unmodified DLC substrates is the profile of these changes. As analysis depth increases (i.e. takeoff angle decreases), for the Ar\textsuperscript{+} ion sputtered DLC the increase in the CO/CN group (and the corresponding decrease in the CC/CH group) begins earlier than for the unmodified DLC. This is a consequence of the decreased thickness of the protein layer on the Ar\textsuperscript{+} ion sputtered DLC compared to the unmodified DLC (see Table \ref{tab:surfacechem}, which allows stronger signals from the groups located closer to the DLC-protein interface at higher takeoff angles.\\

The important role of CO/CN groups in bonding between the DLC surface and protein layer is due to the increased strength of polar interactions compared to the much weaker dispersive interactions experienced by CC/CH groups. The COOH/CONH group, on the other hand, does not change with analysis depth for any of the three samples - implying that it may not have much contribution to bonding between the adsorbed BSA and the surface. This is probably due to the larger size of these groups, making it more difficult for them to change orientation and participate in bonding.\\

\begin{figure}
\centering
\includegraphics{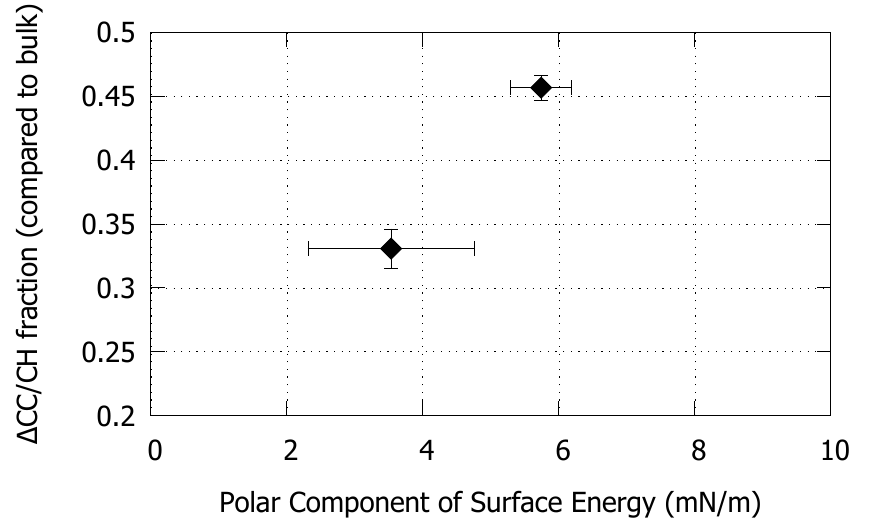}
\textbf{\caption{\label{fig:Fig_7}Graph showing the increase in segregation of the CC/CH group with increasing polar surface energy.}}
\end{figure}

It is particularly interesting to observe the relationship between the segregations observed here, and the surface energies measured. Figure \ref{fig:Fig_7} shows the magnitude of the segregation of the non-polar CC/CH group (in terms of the change in composition between the bulk BSA and the adsorbed BSA at 80$^{\circ}$ takeoff angle) as a function of the polar component of suface energy ($ \gamma^{p}_{S} $). It has previously been shown \cite{Feng03} that the polar component of surface energy is the dominant factor in adsorption of BSA onto solid surfaces - as $ \gamma^{p}_{S} $ increases, adsorption increases. The results here support this, and demonstrate that due to the important role polar groups (CO/CN, COOH/CONH) play in bonding, non-polar CC/CH groups are segregated away from the interface. When the polar component of surface energy is higher, segregation of BSA groups increases - this is shown here by the increase in non-polar group (CC/CH) segregation away from the interface.\\

In addition to analysing the variations in these groups with depth, we can also discuss the significance of their relative magnitudes. The BSA adsorbed onto the unmodified DLC coating displays a CC/CH fraction that is consistently 0.1-0.2 higher than the Ar\textsuperscript{+} ion sputtered DLC coating (see Figure \ref{fig:Fig_6}). This can be attributed to the slightly different thicknesses of the adsorbed protein layers mentioned above. Since we have established that the proportion of CC/CH bonded carbon decreases as we move down the protein layer towards to DLC-protein interface, it follows that the XPS signal from a thicker protein layer will contain a higher proportion of carbon in this configuration. This explains the difference between the baseline CC/CH concentrations of the DLC and Ar\textsuperscript{+} ion sputtered DLC.

\section{Conclusions}
Significant changes in the surface chemistry of the DLC coatings were achieved after modification. Laser graphitisation was seen to increase the amount of CO and C=O groups on the DLC surface, leading to an increase in the polar component of surface energy. Ar\textsuperscript{+} ion sputtering decreased the amount of CO and C=O groups, leading to a corresponding decrease in the polar component of surface energy.\\

Angle Resolved X-ray Photoelectron Spectroscopy (ARXPS) of the adsorbed BSA layer shows that polar groups (i.e. CO/CN) are segregated towards the DLC-protein interface - this is due to their important role in bonding. It is known that the polar component of surface energy affects protein adsorption onto solid surfaces, with a higher $ \gamma^{p}_{S} $ resulting in higher adsorption. This study has shown that an increased $ \gamma^{p}_{S} $ also results in a greater degree of segregation of the polar CO/CN groups involved in bonding, presumably a result of increased net attraction between the solid surface and polar groups at higher polar surface energy.

\section{Acknowledgements}
The authors would like to thank Dr. Colin Doyle for assistance in conducting XPS experiments, and the Photon Factory (University of Auckland) for assistance with laser graphitisation experiments.

\bibliographystyle{model1a-num-names}

\bibliography{Refs}

\FloatBarrier

\begin{flushleft}

\begin{table}
\textbf{\caption{\label{tab:surfacechem}Binding energy components of the modified and unmodified DLC surfaces.}}
\makebox[\linewidth]{
\begin{tabularx}{1.45\textwidth}{ |p{3.3cm}|p{2.4cm}|X|p{2cm}|p{2.8cm}|p{2.8cm}| }
\hline
Component & Binding Energy (eV) & Ref. & DLC & Ar\textsuperscript{+} Ion Sputtered DLC & Laser Graphitised DLC \\
\hline
sp\textsuperscript{3} & 285.4 & \cite{Takabayashi07, Moulder92, Ahmed13} & 41.3$\% (\pm0.1)$ & 15.1$\% (\pm1.6)$ & 27.5$\% (\pm1.8)$\\ 
sp\textsuperscript{2} & 284.8 & \cite{Takabayashi07, Moulder92, Ahmed13} & 53.4$\% (\pm0.9)$ & 79.4$\% (\pm1.9)$ & 51.3$\% (\pm1.2)$\\ 
CO & 286.7 & \cite{Moulder92, Xue12} & 3.8$\% (\pm0.6)$  & 3.6$\% (\pm0.3)$ & 15.7$\% (\pm0.3)$\\ 
C=O & 288.6 & \cite{Moulder92, Xue12} & 1.6$\% (\pm0.2)$  & 1.9$\% (\pm0.02)$ & 5.6$\% (\pm0.4)$\\ 
sp\textsuperscript{3}/sp\textsuperscript{2} ratio & & & 0.77$(\pm0.01)$  & 0.19$(\pm0.02)$ & 0.53$(\pm0.05)$\\
BSA Layer Thickness & & & 39.0\AA{} $(\pm2.3)$  & 29.8\AA{} $(\pm2.1)$ & -\\
\hline
\end{tabularx}
}
\end{table}

\begin{table}
\textbf{\caption{\label{tab:surface-energy}Surface energy of modified and unmodified DLC surfaces.}}
\makebox[\linewidth]{
\begin{tabularx}{1.25\columnwidth}{ |p{5.5cm}|l|X|X| }
\hline
Surface Energy Component & DLC & Ar\textsuperscript{+} Ion Sputtered DLC & Laser Graphitised DLC \\
\hline
Dispersive Surface Energy (mN/m) & 41.6 $\pm$0.7 & 41.4 $\pm$1.3 & 45.0 $\pm$1.0\\ 
Polar Surface Energy (mN/m) & 5.7 $\pm$0.5 & 3.5 $\pm$1.2 & 13.6 $\pm$0.4\\ 
Total Surface Energy (mN/m) & 47.4 $\pm$0.8  & 44.9 $\pm$2.1 & 58.5 $\pm$0.7\\ 
\hline
\end{tabularx}
}
\end{table}

\begin{table}
\textbf{\caption{\label{tab:BSA-standard}Binding energy components of the BSA standard, showing CC/CH, CO/CN, and COOH/CONH components.}}
\begin{tabularx}{\columnwidth}{ |X|l|l| }
\hline
Component & Position (Binding Energy, eV) & Area$\%$ \\
\hline
CC/CH & 285.4 & 42\\
CO/CN & 286.7 & 28\\
COOH/ CONH & 288.6 & 30\\
\hline
\end{tabularx}
\end{table}

\end{flushleft}

\end{document}